\documentclass[twocolumn]{aastex63}

\newcommand{\y}[1]{{\textcolor{red}{#1}}}

\shorttitle{The origin of host mass-step}
\shortauthors{Chung et al.}

\begin{document}

\title{On the root cause of the host `mass-step' in the Hubble residuals of type Ia supernovae}

\correspondingauthor{Chul Chung, Young-Wook Lee}
\email{chulchung@yonsei.ac.kr, ywlee2@yonsei.ac.kr}

\author[0000-0001-6812-4542]{Chul Chung}
\affil{Department of Astronomy \& Center for Galaxy Evolution Research, Yonsei University, Seoul 03722, Republic of Korea}
\affil{Equal First Authors}

\author[0000-0002-1842-4325]{Suk-Jin Yoon}
\affil{Department of Astronomy \& Center for Galaxy Evolution Research, Yonsei University, Seoul 03722, Republic of Korea}
\affil{Equal First Authors}

\author{Seunghyun Park}
\affil{Department of Astronomy \& Center for Galaxy Evolution Research, Yonsei University, Seoul 03722, Republic of Korea}

\author{Seunghyeon An}
\affil{Department of Astronomy \& Center for Galaxy Evolution Research, Yonsei University, Seoul 03722, Republic of Korea}

\author{Junhyuk Son}
\affil{Department of Astronomy \& Center for Galaxy Evolution Research, Yonsei University, Seoul 03722, Republic of Korea}

\author[0000-0001-5966-5072]{Hyejeon Cho}
\affil{Department of Astronomy \& Center for Galaxy Evolution Research, Yonsei University, Seoul 03722, Republic of Korea}

\author[0000-0002-2210-1238]{Young-Wook Lee}
\affil{Department of Astronomy \& Center for Galaxy Evolution Research, Yonsei University, Seoul 03722, Republic of Korea}

\begin{abstract}

It is well established that the Hubble residuals of type Ia supernovae (SNe Ia) show the luminosity step with respect to their host galaxy stellar masses. 
This `mass-step' is taken as an additional correction factor for the SN Ia luminosity standardization. 
Here we investigate the root cause of the mass-step and propose that the bimodal nature of the host {\it age} distribution is responsible for the step. 
In particular, by using the empirical {\it nonlinear} mass-to-age relation of local galaxies, we convert the mass function of SN Ia hosts to their age distribution.
We find that the age distribution shows clear bimodality: a younger ($<$\,6\,Gyr) group with lower mass ($\sim$\,$10^{9.5}$\,${\rm M}_{\rm sun}$) and an older ($>$\,6\,Gyr) group with higher mass ($\sim$\,$10^{10.5}$\,${\rm M}_{\rm sun}$).
On the Hubble residual versus host mass plane, the two groups create the mass-step at $\sim$\,$10^{10}$ ${\rm M}_{\rm sun}$.
This leads us to conclude that the host galaxy mass-step can be attributed to the bimodal age distribution in relation to a nonlinear relation between galaxy mass and age.
We suggest that the mass-step is another manifestation of the old `red sequence' and the young `blue cloud' observed in the galactic color--magnitude diagram.

\end{abstract}


\section{Introduction}

The type Ia supernovae (SNe Ia) are currently  the most important standard candles in observational cosmology beyond the local universe because they allow the direct measurement of the expansion rate of the universe \citep{1998AJ....116.1009R, 1998ApJ...507...46S, 1999ApJ...517..565P}.
Contrary to the initial assumption that SNe Ia are reliable ``standardizable candles'', studies over the past decade have revealed that the brightnesses of SNe Ia, even after the luminosity standardization, in fact vary depending on their host galaxy properties.
The host properties suggested thus far include host stellar mass \citep[e.g.,][]{2010ApJ...715..743K, 2010MNRAS.406..782S}, local star formation rate  \citep[e.g.,][]{2013A&A...560A..66R, 2015ApJ...802...20R, 2020A&A...644A.176R}, metallicity \citep[e.g.,][]{2003MNRAS.340.1057S, 2010MNRAS.406..782S}, dust extinction \citep[e.g.,][]{2018ApJ...859..101S, 2021ApJ...909...26B}, and (local/global) progenitor age \citep[e.g.,][]{2009ApJ...691..661H, 2011ApJ...740...92G, 2016ApJS..223....7K, 2020ApJ...903...22L, 2022MNRAS.tmp.2647L, 2021MNRAS.503L..33Z, 2023arXiv230315267W}. 
In particular, recent studies of SN cosmology essentially include the host mass correction in the luminosity standardization of SNe Ia.

\begin{figure*}
\centering
\includegraphics[angle=0,scale=0.67]{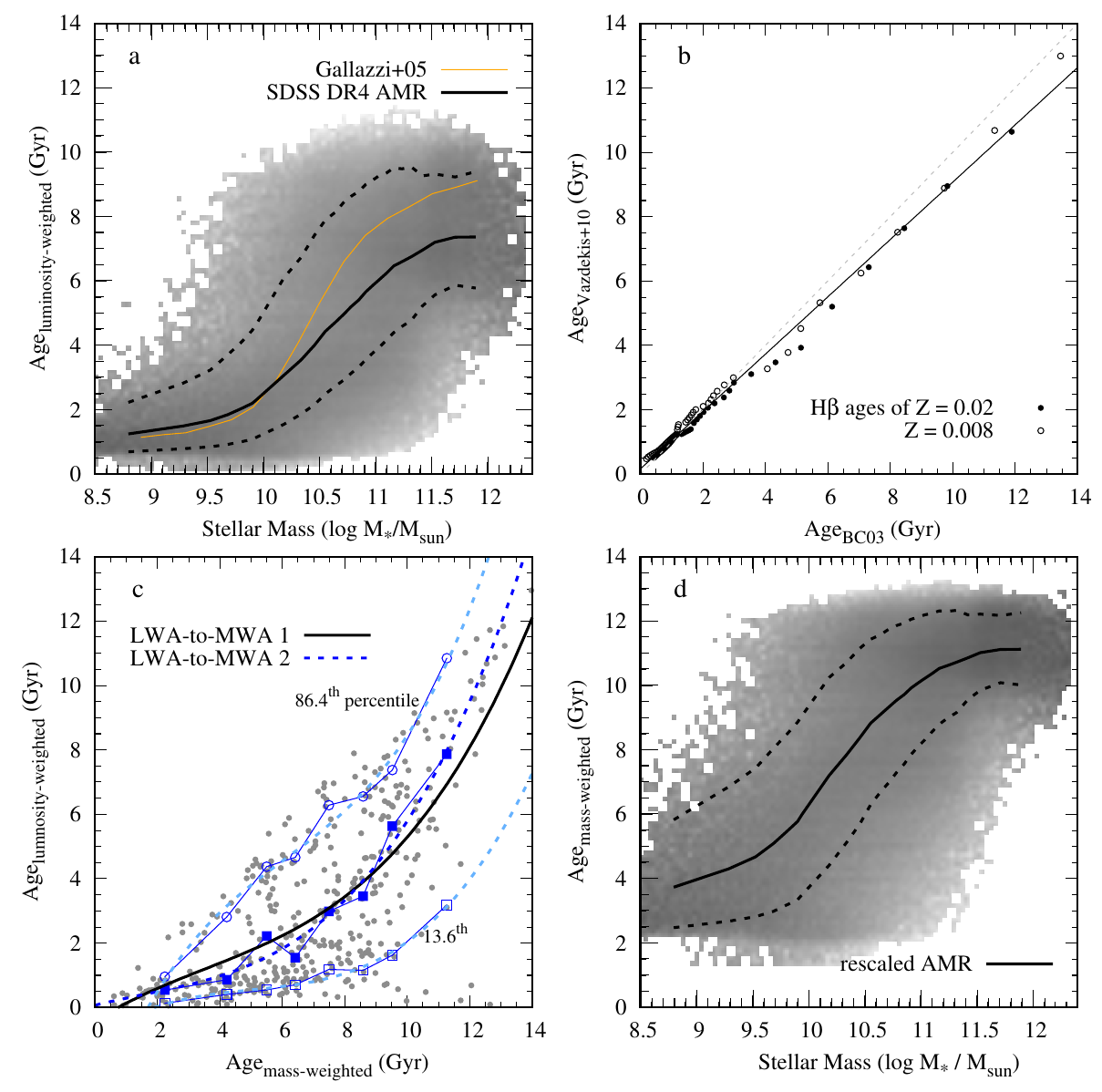}
\caption{
($a$) The SDSS DR4 version of G05 age--mass relation.
The logarithmic normalized density map of $\sim$420,000 galaxies in SDSS DR4 is shown in grayscale.
The black solid line traces the median age value of $\sim$5000 galaxies across the entire mass range.
The black dashed lines correspond to the $\pm 34.1$\% scatter from the median age.
The empirical relation from G05 based on SDSS DR2 is depicted by the orange line.
($b$) Comparison of H$\beta$-derived ages between \citet{2003MNRAS.344.1000B} and \citet{2010MNRAS.404.1639V}. 
The model is for Z = 0.008 and Z = 0.02.
The black solid line indicates the linear regression between the two age scales.
($c$) The relationship between LWA and MWA in the CALIFA galaxies \citep{2017MNRAS.471.3727D}.
The black solid line indicates a third-order polynomial regression that closely follows the ages of the CALIFA galaxies.
The blue dashed line represents the third-order polynomial regression obtained from the median values of 8 bins having $\sim$50 galaxies each (blue squares). 
The 13.6th and 86.4th percentile values are represented as blue open squares and open circles, respectively. 
Following these points, light blue dashed lines then illustrate third-order polynomial regressions for each percentile.
($d$) The rescaled AMR based on the CALIFA galaxies.
By converting the ages between population models and the LWA-to-MWA scale, our rescaled AMR for the MWA is represented by the black solid line.
The grayscale map represents the logarithmic normalized density of the rescaled SDSS DR4 galaxies.
}
\label{f1}
\end{figure*}

\begin{figure*}
\centering
\includegraphics[angle=0,scale=0.75]{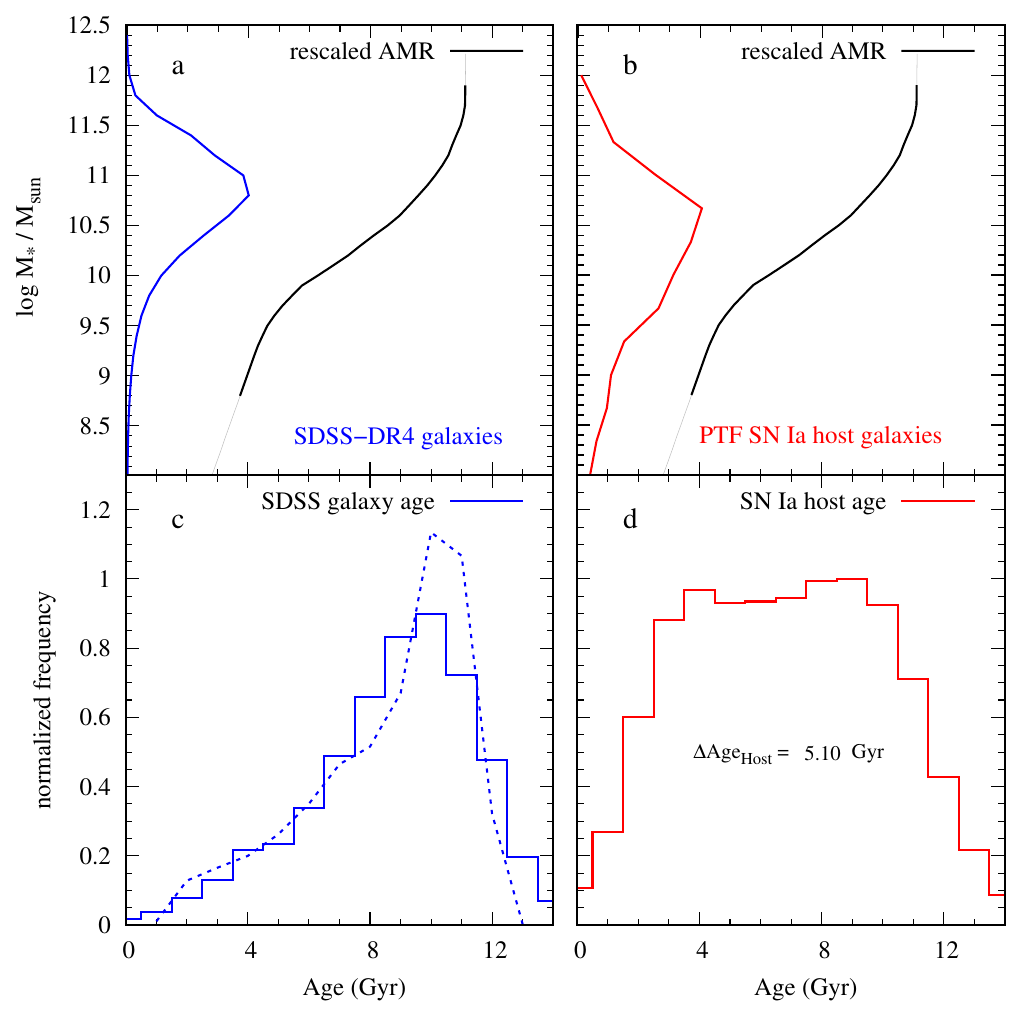}
\caption{
The projection simulations based on ($a$) the mass function of $\sim$420,000 SDSS DR4 galaxies (the updated version of G05) and ($b$) the PTF host mass distribution \citep[][]{2014MNRAS.438.1391P}.
The thin gray lines extending along the AMR represent extrapolations for galaxies of both low and high masses.
The resulting age distributions projected from the mass distributions are shown in panels ($c$, $d$).
The age scatter derived from Figure~\ref{f1}($d$) is assumed for the projected age distributions.
Galaxies that lie significantly outside the AMR are obtained through extrapolations at both ends.
The measured age distribution of SDSS DR4 galaxies is presented as a dashed blue line in panel ($c$).
The blue histograms in panel ($c$) have been normalized such that their areas are equivalent.
Both the host age of SN Ia and the age of the SDSS DR4 galaxies represent global ages for each respective galaxy.
The difference in the mean age between young and old groups for the simulated distribution of hosts is analyzed by the GMM test and shown in panel ($d$).
}
\label{f2}
\end{figure*}

While the flux-limited host mass distribution is unimodal with slight skewness toward lower masses, the standardized brightnesses of SNe Ia show a significant difference below/above a certain host mass ($\sim 10^{10} {\rm M}_{\rm sun}$). 
Currently, the luminosity corrections of SNe Ia follow this host `mass-step' \citep[e.g.,][]{2014A&A...568A..22B, 2020MNRAS.494.4426S, 2021ApJ...923..237J}. 
However, given that the mass of the host galaxy has no direct relevance to the nature of SNe Ia, previous studies have suggested that this phenomenon stems from either age/metallicity \citep[e.g.,][]{2009ApJ...691..661H, 2010MNRAS.406..782S} of underlying stellar populations or dust properties of SN Ia hosts \citep[e.g.,][]{2021ApJ...909...26B, 2021ApJ...923..237J}.
Recently, \citet{2023MNRAS.520.6214W} proposed that the dust extinction parameter $R_{\rm V}$, which varies with the host age, could be the driving factor behind the host mass-step.
Since the stellar mass of a galaxy is correlated with the population age \citep[e.g.,][]{2005MNRAS.362...41G, 2016ApJS..223....7K}, the host mass-step may be an outcome of the progenitor age effect on the standardized brightnesses of SNe Ia \citep{2020ApJ...889....8K, 2020ApJ...903...22L, 2022MNRAS.tmp.2647L} rather than being mainly influenced by dust \citep[see e.g.,][]{2020ApJ...901..143U, 2021ApJ...923..197P, 2022MNRAS.517.2360T}.
Indeed, \citet{2014MNRAS.445.1898C} have demonstrated how the host mass-step could be driven by progenitor age, although they did not proceed to the simulation for the Hubble residual (HR) versus host mass diagram.
Identifying the root cause of the mass-step is crucial as each host galaxy property evolves differently with redshift and, therefore, has a different impact on cosmology.
Within the redshift range most relevant to SN cosmology ($0 < z < 1$), the variations in host mass or star formation rate are either negligible or relatively weak, producing only an insignificant or limited effect on cosmology \citep{2018ApJ...859..101S, 2020A&A...644A.176R}. 
On the other hand, the variation in mean population age is significant, causing a critical impact on cosmology \citep{2022MNRAS.tmp.2647L}.

 This paper aims to investigate the primary cause of the host galaxy mass-step, instead of a linear relationship, by employing the empirical correlation between galaxy mass and population age.
 This study will specifically focus on the host age among the various parameters that have been suggested to be associated with the host mass-step.
We demonstrate here that the mass-step observed in SN Ia host galaxies is a natural consequence of a simple reflection from the nonlinear relation between galaxy mass and mean population age, which is also witnessed as old `red sequence' and young `blue cloud' in the galaxy color--magnitude diagram, as well documented in the literature.

\section{The Empirical Age--Mass Relation of galaxies}

\citet[][hereafter G05]{2005MNRAS.362...41G} found the nonlinenar empirical age--mass relation (AMR) of galaxies by estimating the luminosity-weighted age (LWA) through ${\rm D4000}$ and Balmer absorptions, based on the SDSS DR2.
Subsequently, the G05 AMR was validated by \citet{2008MNRAS.383.1439G,2021MNRAS.502.4457G} and \citet{2017MNRAS.468.1902Z}.
To update the G05 AMR, we apply the same methodology of G05 to analyze $\sim$\,420,000 galaxies\footnote{https://wwwmpa.mpa-garching.mpg.de/SDSS/DR4/} from SDSS DR4.
The choice of SDSS DR4 is because of its availability of the spectroscopically measured age, which is considered as an accurate method of age-dating galaxies. 
This sample of galaxies provides a good approximation for the AMR required in this analysis as SNe Ia arise in all types of galaxies. 
In Figure~\ref{f1}($a$), we present our newly derived AMR for SDSS DR4 galaxies. 
The age difference between massive and low-mass galaxies decreased by $\sim$\,10\,\% compared to the G05 AMR for SDSS DR2 (orange line). 
This difference is attributed to the mass offset between G05 (SDSS DR2) and SDSS DR4, as well as the younger galaxies of SDSS DR4 within massive galaxies around $z \sim 0.3$.
As demonstrated, sample selection influences the resultant AMR. 
Nonetheless, as elaborated in subsequent discussions, the essence of the empirical AMR resides in its nonlinear relationship. 
Hence, we choose to employ this AMR for diverse analyses of SN Ia hosts throughout this study.

In this study, we adopt the mass-weighted age (MWA) \citep[][]{2020ApJ...903...22L} for host galaxies. 
Compared to the LWA, the MWA more accurately represents the mean age of stellar populations in a galaxy especially when recently formed young stars are present.
Moreover, given the significant influence of the star formation history and the delay time distribution on the SN progenitor age distribution, the MWA is considered more suitable for investigating the correlation between mass and age of hosts.
To convert from an LWA-based AMR to an MWA-based one, we use data from the CALIFA survey \citep{2017MNRAS.471.3727D}, which provides both LWA and MWA measurements.
It is noteworthy that the LWA of our AMR is based on the \citet[][hereafter BC03]{2003MNRAS.344.1000B} model, while the CALIFA survey's LWA is based on the \citet[][hereafter V10]{2010MNRAS.404.1639V} model.
To establish a correlation between ages acquired from the BC03 and V10 models, we obtain simple stellar population equivalent ages based on the H$\beta$ absorption index (for $Z = 0.008$ and $Z = 0.02$) in Figure~\ref{f1}($b$) and derive a formula, ${\rm Age}_{\rm V10}=0.890 \times {\rm Age}_{\rm BC03} + 0.167$ (solid line).
Using this formula, we convert the updated G05 AMR to a rescaled AMR with the age scale of the CALIFA survey.

\begin{figure*}
\centering
\includegraphics[angle=0,scale=0.75]{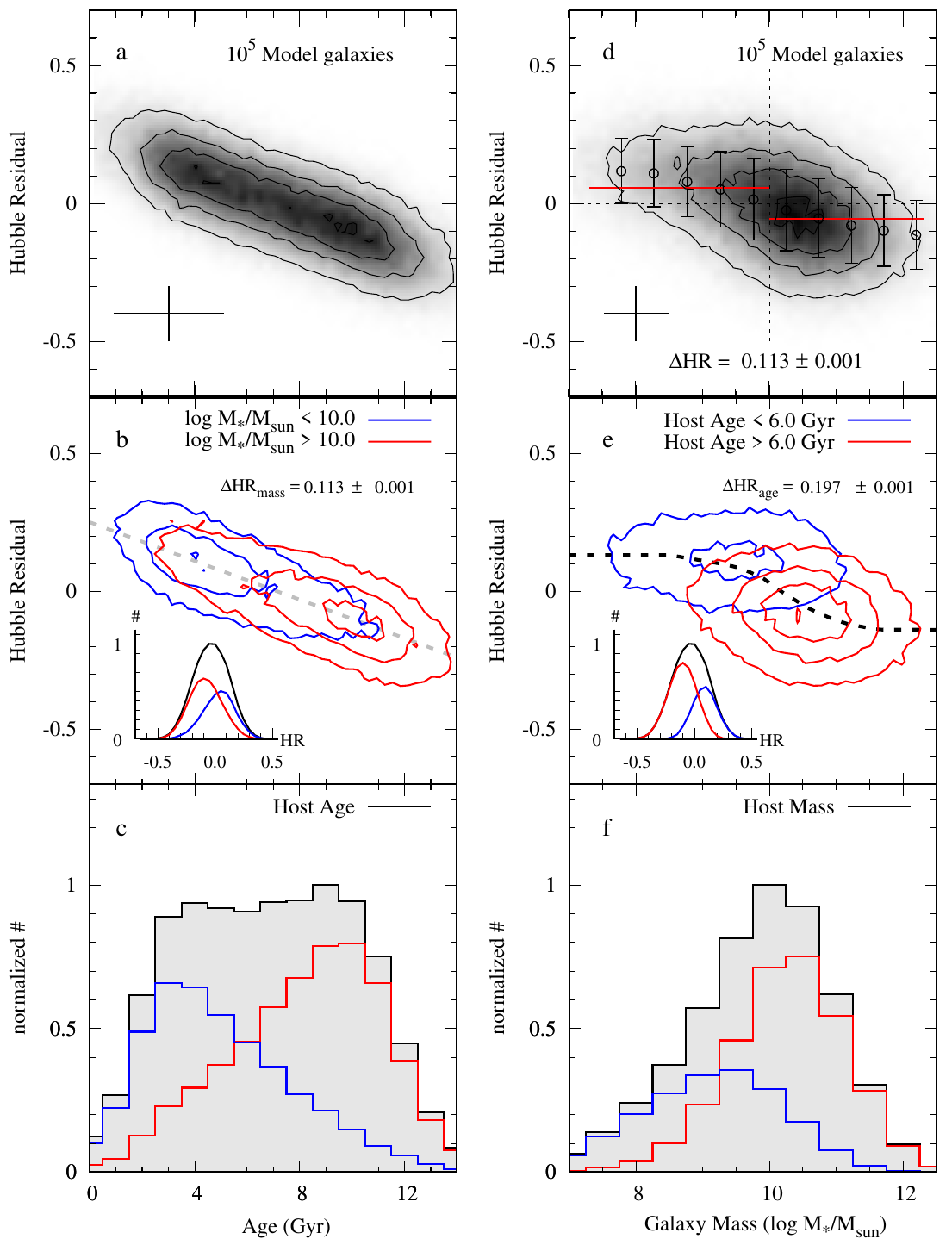}
\caption{
Reproducing the mass-step from the host age and host mass of SN Ia.
Top rows: A probability density map is shown based on $10^5$ mock galaxies, with ($a$) and ($d$) panels depicting the host age versus HR and the host mass versus HR simulations, respectively.
The typical errors applied for the simulation are presented at the bottom left.
Open circles in ($d$) represent binned average HRs in host mass bins of width 0.5 dex, while thick red lines denote the mean HRs of host masses above and below the $\log ({\rm M}_{*}/{\rm M}_{\rm sun}) = 10.0$ threshold (vertical dashed line).
The host mass-step ($\Delta$HR) is indicated at the bottom.
Middle rows: The contours above and below the host mass ($\log {\rm M}_{*}/{\rm M}_{\rm sun} = 10.0$) and the host age ($6.0$~Gyr) thresholds are depicted in panels ($b$) and ($e$), respectively.
The gray dashed line on ($b$) corresponds to the slope of the age--HR relationship ($\Delta{\rm HR}/\Delta{\rm age}=-0.035$ mag/Gyr), and the empirical AMR within the host mass versus HR plane is represented by the black dashed curve on ($e$).
The insets display histograms of the HR distribution for each subgroup, with the identical color codes as applied in the contours.
Bottom rows: The host age and the host mass distributions of simulated galaxies.
The distribution of host ages shows bimodality, while the distribution of host masses is unimodal.
The red and blue histograms in each plot correspond to the respective color contours presented in the middle rows.
}
\label{f3}
\end{figure*}

\begin{figure*}
\centering
\includegraphics[angle=0,scale=0.69]{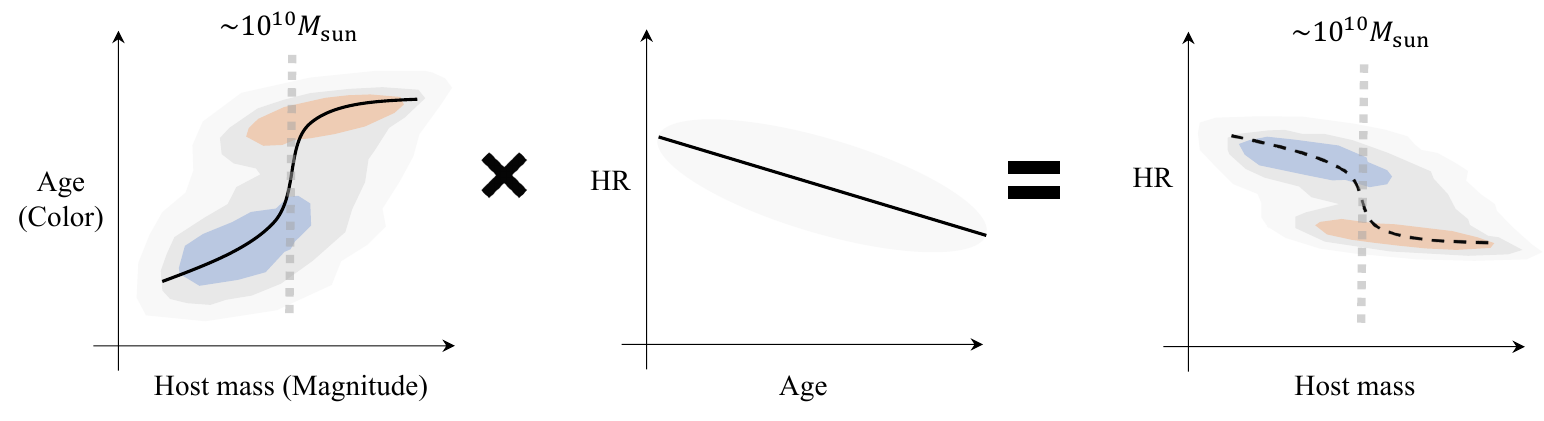}
\caption{
A schematic diagram explaining the origin of the host galaxy mass-step using two relationships of the galaxy age versus galaxy mass and HR versus host age. 
Left panel: A typical galaxy color--magnitude diagram is presented with the highlighted red sequence and blue cloud regions. 
The black solid line represents the empirical relation between galaxy age and mass utilized in this study. 
Middle panel: It is assumed that the HR is linearly correlated to the host age. 
Right panel: The convolution of the host age to the host mass results in the HR with respect to the host mass diagram. 
The schematic diagrams contain gray dashed lines that approximately represent $10^{10} {\rm M}_{\rm sun}$.
The predicted sequence of host galaxies on the HR and host mass diagram is denoted by the black dashed line. 
The fundamental structure of the host galaxy distribution, which takes the form of a step function, persists in the HR versus host mass plane and can be discerned as the host galaxy mass-step.
}
\label{f4}
\end{figure*}

Figure~\ref{f1}($c$) shows the relationship between MWA and LWA in CALIFA galaxies. 
We note that the galaxies with MWAs below $\sim$9~Gyr are mostly confined to LWAs of $<2$~Gyr.
This illustrates the effect of young stellar populations on the LWA. 
To obtain the LWA-to-MWA conversion relationship, we perform polynomial regressions between LWAs and MWAs and choose to use the third-order polynomial (black solid line, LWA-to-MWA conversion 1), which yields the minimum value of reduced $\chi^2$.\footnote{To obtain the reduced $\chi^2$ value, we assigned equal weight to each age scale, given the limited information available on the error bars for both MWA and LWA. Among the 12 regressions tested (2nd to 7th order polynomials) for both MWA and LWA, the 3rd order polynomial shown in Figure~\ref{f1}($c$) had the lowest reduced $\chi^2$ value. However, it is important to note that this minimum value does not signify that the regression is the optimal fit, but rather underscores its superior fit within the range of 2nd to 7th order polynomials.}
As an alternative approach, we divide the samples into eight groups, each consisting of $\sim$50 galaxies, and obtain the median values for each group (blue squares in the figure). 
Utilizing these data points, we derive a third-order polynomial regression (blue dashed line, LWA-to-MWA conversion 2).
To capture the full scatter in both conversions, we further integrate conversion errors between two age scales using the 13.6th and 86.4th percentile sample distributions  of LWA-to-MWA conversion 2.
Note that our simulations in Figures~\ref{f2}-\ref{f4} do not incorporate these LWA-to-MWA conversion uncertainties, as our primary focus is to illustrate typical age scale conversions. 
A detailed discussion on LWA-to-MWA conversion errors is provided in the Appendix.
In Figure~\ref{f1}($d$), we present the MWA-based AMR (black solid line), whose shape remains almost the same as the LWA-based AMR.
Regardless of the selected LWA-to-MWA conversion or the inclusion of age conversion errors, the resulting AMRs consistently exhibit a highly nonlinear `S'-shape with a quasi-inflection point between young and old ages.
This nonlinearity recalls the color$-$magnitude (mass) relation in the diagnostic diagrams of galaxy evolution \citep[e.g.,][]{2007ApJ...665..265F}.
It is noteworthy that this nonlinear relationship is also observed in the host age--mass relationship of SN Ia hosts as observed by \citet{2011ApJ...740...92G}. 
This further affirms that our AMR derived from SDSS DR4 serves as a robust approximation for a wide mass range of galaxy samples.

\section{The bimodal age distribution of galaxies as the cause of the host mass-step}

The observed AMR shows a nonlinear pattern that reflects the average relationship between galaxies' age and mass.
In order to investigate the age distribution of galaxies based on this empirical AMR, we conduct the nonlinear projection simulations.\footnote{Detailed explanation on the nonlinear projection simulation relevant to this work can be found in \citet{2006Sci...311.1129Y}, \citet{2016ApJ...818..201C}, and \citet{2019ApJS..240....2L}.}
Figure~\ref{f2}($a$, $b$) shows two simulations using the observed mass distributions of ($a$) the SDSS DR4 galaxies and ($b$) the SN Ia host galaxies from the Paloma Transient Factory \citep[PTF,][]{2014MNRAS.438.1391P}.
The SN Ia host galaxies show a broader mass distribution with a higher fraction at the low-mass tail.
This is attributed to SNe Ia arising more frequently in low-mass star-forming galaxies.

In Figure~\ref{f2}($c$, $d$), we convert the two mass distributions into the age distributions using the nonlinear AMR.
Changes in the slopes of the empirical AMR can affect the concentration and dispersion of the age distribution of galaxies.
Additionally, both the location of the peak and the degree of spread in a mass distribution play a role in determining the age distributions, resulting in high skewness for SDSS galaxies and strong bimodality for SN Ia hosts.
As a mass distribution shifts towards lower mass or becomes wider with a greater proportion of low-mass galaxies, the age distributions become more strongly bimodal.
In this regard, applying a completeness correction to the mass distribution would result in an age distribution exhibiting increased numbers of younger galaxies.
However, considering the variations within the blue cloud region where low-mass galaxies predominantly reside, the overall AMR shape is anticipated to remain relatively stable.
We note that, despite the AMR does not cover the entire age range, the simulated ages exhibit a distribution across the entire age span. 
This is due to a mild extrapolation of the rescaled AMR and the Monte Carlo simulation to introduce a random age scatter in the projection simulation. 
While a small number of galaxies in the young age region may appear unphysical, their contribution to the overall age distribution is negligible.

In Figure~\ref{f2}($c$), we present a validation of our projection simulation result by comparing the simulated age distribution of SDSS DR4 galaxies (blue solid histogram) with their measured age distribution (blue dashed line).
The simulated age distribution exhibits a peak at an older age and a skewed tail towards younger ages, which closely resembles the measured age distribution.
Therefore, in Figure~\ref{f2}($d$), our simulated age distribution for the host galaxies can be considered as a reasonable representation to their actual age distribution.

\begin{figure*}
\centering
\includegraphics[angle=-90,scale=0.70]{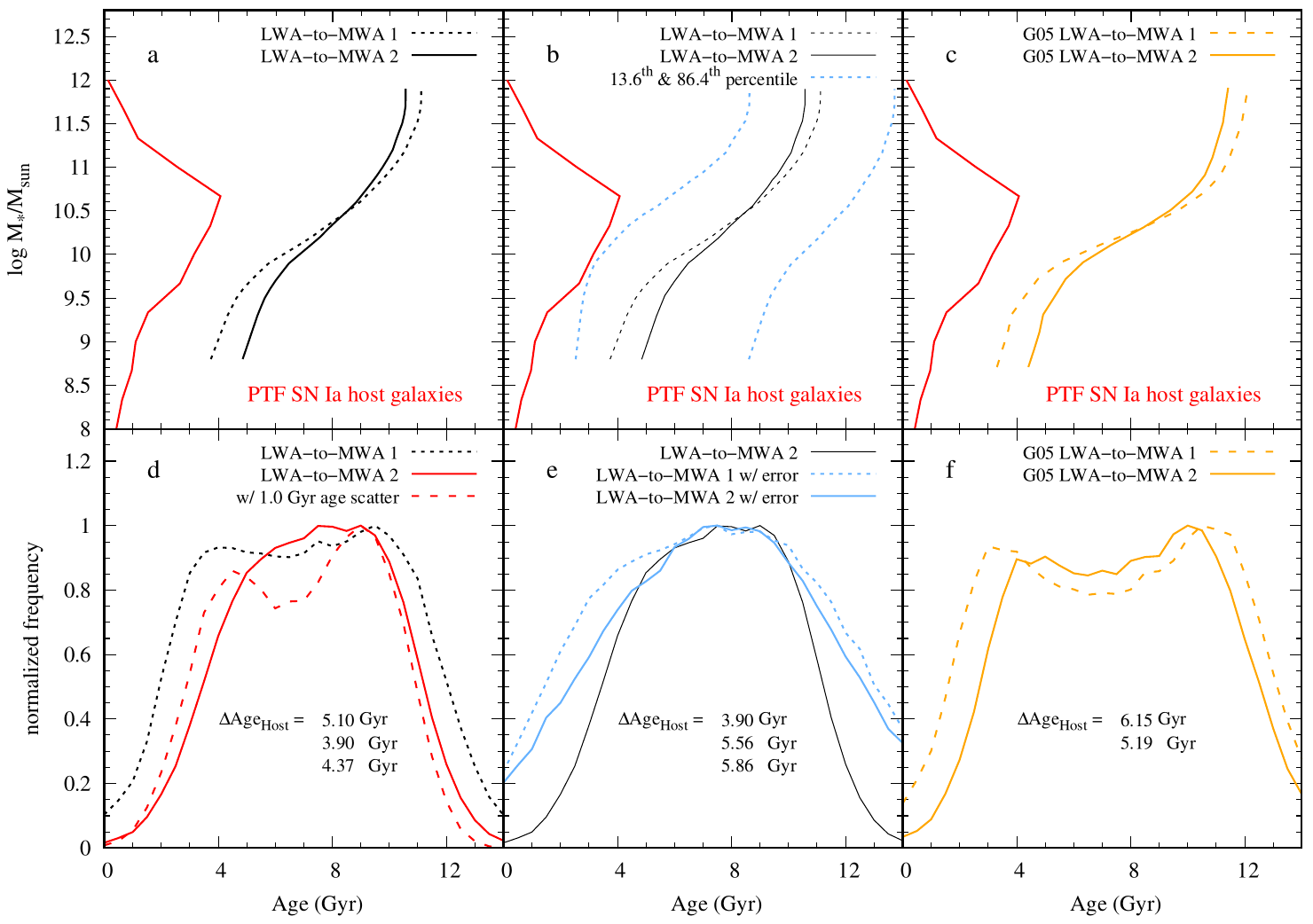}
\caption{
Projection experiments for different assumptions on AMRs and the age uncertainty.
Upper row: Panel ($a$) presents a comparison between the rescaled AMR obtained from LWA-to-MWA conversion 1 (short dashed line) used throughout the manuscript and the AMR derived from the LWA-to-MWA conversion 2 shown in Figure~\ref{f1}($d$) (solid line). 
Panel (b) shows an additional pair of AMRs resulting from 13.6th and 86.4th percentile conversions (light blue short dashed lines) of LWA-to-MWA conversion 2.
The AMRs for the G05, after being rescaled under the same conversions as panel ($a$), are represented by the orange long-dashed and solid lines in panel ($c$). 
The mass distribution employed for the projection simulation is identical to that in Figure~\ref{f2}($b$).
Lower row: The resulting age distributions, corresponding to the AMRs mentioned above, are displayed with the line type matching those in the upper row. 
The black short dashed histogram in panel ($d$) is the same age distribution in Figure~\ref{f2}($d$). 
The red dashed histogram illustrates the age distribution using the age scatter of 1.0~Gyr.
The age distributions, resulting from LWA-to-MWA conversions 1 and 2, which include extreme outliers in the age conversion as depicted in panel (b), are presented in panel (e). 
The age distributions maintain the line types of LWA-to-MWA conversion 1 and 2 from panel (b) with the light blue colors.
Panel ($f$) displays the age distributions based on the G05 AMRs obtained using different conversions.
Irrespective of the AMRs applied, the projection simulation consistently yields bimodal age distributions.
}

\label{f1a}
\end{figure*}

To quantify the degree of bimodality in age distributions resulting from the projection simulation, we execute the Gaussian Mixture Modeling (GMM) test \citep{2010ApJ...718.1266M} with different variances, using $10^4$ model galaxies chosen randomly.
The GMM test yields $D$-values\footnote{The $D$-value is defined as ${D \equiv | \mu_1 - \mu_2 |/ [(\sigma^2_1 + \sigma_2^2)/2]^{1/2}}$, where $\mu_1$ and $\mu_2$ are means for two distributions with $\sigma_1$ and $\sigma_2$ for standard deviations, respectively.} and the fraction between two groups, which is a useful diagnostic for the separation between two distributions: a $D$-value of at least 2.0 is necessary for a clear separation between two groups \citep{1994AJ....108.2348A}.
Specifically, the $D$-value for SDSS DR4 galaxies is 2.41 but the distribution exhibits a dominant fraction of older age groups ($80.9\pm2.2$\%).
In contrast, for the hosts, the GMM test shows a distinct separation between the two age groups, with a $D$-value of 2.49, along with a comparable fraction of young ($37.0 \pm 2.4$\%) and old galaxies ($63.0 \pm 2.4$\%).
Note that employing the rescaled G05 AMR results in an even more pronounced bimodal age distribution ($D$-value = 2.88) with a similar fraction of two groups and a slightly larger mean age difference (see Appendix for the age distributions with different input ingredients and parameters).

In Figure~\ref{f3}, we perform Monte Carlo simulations with $10^5$ mock galaxies to examine whether the bimodal age distribution of SN Ia hosts can account for the observed host mass-step.
For the age--HR simulation, we take into account age scatters derived from Figure~\ref{f1}($d$) to the projected age while using a mass distribution from PTF. 
For the mass–HR simulation, each age of a mock galaxy is fixed while incorporating mass scatter derived from the SDSS DR4 data into the mass of a galaxy. 
In all simulations, the HR scatter is consistently maintained at a $1\sigma$ value of 0.1~mag  \citep{2014MNRAS.438.1391P}.
Firstly, in the left panels, we examine the luminosity difference of SNe Ia as a function of the host age. 
Figure~\ref{f3}($a$) displays the distribution of mock galaxies in the host age versus HR plane.
As noted by \citet{2020ApJ...903...22L}, the potential effect of a nonlinear age--HR is not significant, and we adopt a linear relation\footnote{Per-object SN analysis concerning host age and HR has been presented in \citet{2016ApJS..223....7K, 2020ApJ...889....8K} and \citet{2020ApJ...903...22L}. To derive collective trends for HR and host age, we have incorporated the linear relationship as established in literature.} with a $\Delta {\rm HR}/\Delta {\rm age}$ slope of $-0.035$~mag/Gyr \citep[see e.g.,][]{2021MNRAS.503L..33Z, 2022MNRAS.tmp.2647L}.
In Figure~\ref{f3}($b$), we divide the mock galaxies into two groups: those with the mass above (red) and below (blue) the $10^{10} {\rm M}_{\rm sun}$ threshold. 
Although there is a considerable overlap between the two mass groups, they exhibit a noticeable difference of $\Delta {\rm HR} = 0.113$~mag. 
Figure~\ref{f3}($c$) presents the bimodal distribution of host ages, which consists of the age distributions of the two mass groups (see also Figure~\ref{f2}($d$)).

Secondly, in the right panels of Figure~\ref{f3}, we examine the luminosity difference of SNe Ia as a function of host mass (i.e., the mass-step). 
In Figure~\ref{f3}($d$), we generate the host mass versus HR distribution using the empirical AMR (as in Figure~\ref{f1}) and the age$-$HR relation (as in Figure~\ref{f3}(a)).
On closer inspection, the HRs with respect to the host mass exhibit a step function around $10^{10} {\rm M}_{\rm sun}$.
To illustrate this mass-step, we follow the approach adopted in the previous studies reporting the host mass-step, and present the binned HRs (open circles) and the mean values above/below $10^{10} {\rm M}_{\rm sun}$ (red straight lines).
This clearly demonstrates the 0.113~mag step in the HR.
Similar to Figure~\ref{f3}($b$), in Figure~\ref{f3}($e$), we divide the mock galaxies into two groups but based on a host age threshold of 6.0~Gyr.
Interestingly, galaxies divided at 6.0~Gyr are located at different regions in the HR versus host mass plane. 
The galaxies with relatively younger ages mostly fall within the upper-left region, while those with older ages are in the bottom-right region. 
The bimodal age distribution in the host mass-step diagram is attributed to a nonlinear mass--HR relationship stemming from the nonlinear mass--MWA relation, as shown by the black dashed line. 
Despite the substantial overlap in host mass between these two age groups (as shown in Figure~\ref{f3}($f$)), the grouping by host age shows a more pronounced step in the HR (0.197~mag) than that by host mass (0.113~mag).
This suggests that the mass-step is a manifestation of the age-driven HR difference in the host mass domain. 

\section{Discussion}
We have demonstrated how the nonlinear empirical AMR can project a unimodal host mass distribution onto a bimodal host age distribution, thereby providing an explanation for the host mass-step of SNe Ia.
The highly nonlinear empirical AMR is a different manifestation of the distribution of galaxies observed in the color--magnitude diagram.
The distribution of dereddened local galaxies is primarily concentrated in the red sequence (old passive) and blue cloud regions (young star-forming) of the color--magnitude diagram.
This feature has been used to infer the formation and evolutionary history of galaxies at various redshifts \citep[e.g.,][]{2007ApJ...665..265F, 2007ApJS..173..293W, 2013ApJ...777...18M, 2014MNRAS.440..889S}.
In particular, the dereddened UV-optical color, known for its sensitivity to the Balmer/$4000 \AA$ break \citep{2007ApJ...655...51W}, has been employed to classify galaxies into the red sequence or blue cloud categories. 
Given that the Balmer absorption and the D4000 break are well-established age indicators that are relatively insensitive to metallicity \citep[e.g.,][]{2005MNRAS.362...41G, 2013ApJS..204....3C}, the dereddened UV-optical related colors from the $GALEX$ and SDSS survey serve as reliable age / star formation rate indicators of galaxies in the galaxy color--magnitude diagram \citep[e.g.,][]{2007ApJS..173..293W, 2014MNRAS.440..889S}.
As the SN Ia host galaxies can be considered a subset of a larger population of galaxies, local SN Ia host galaxies are also expected to be observed in the same regions as local galaxies in the galaxy color--magnitude diagram.
Therefore, the bimodal age distribution of host galaxies would be a natural consequence of the division of the red sequence and blue cloud in the galaxy color--magnitude diagram.
The empirical AMR and its nonlinearity, which we have utilized to convert host mass to age, would have also been derived from the galaxy color--magnitude relation.

Figure~\ref{f4} is a schematic diagram illustrating the convolution process by which the color--magnitude (or color--mass) diagram of local galaxies is transformed into the HR versus host mass plane.
In the left panel, a typical color--magnitude diagram of local galaxies is depicted, displaying the presence of a nonlinear relationship between the two parameters.
The middle panel shows a correlation between the ages of host galaxies and their corresponding SN Ia HRs. 
Notably, the observed host mass-step in the SN Ia host galaxies is the result of a simple convolution of these two relations, as illustrated in the right panel.
In this regard, the host mass-step, which is observed to occur at approximately $10^{10} {\rm M}_{\rm sun}$ can also be naturally explained by the transition mass between the red sequence and blue cloud regions in the galaxy color--magnitude (mass) diagram \citep[e.g.,][]{2007ApJ...665..265F}.
This is also in line with the findings of \citet{2018A&A...615A..68R}, who reported a significant local $U-V$ color step that is comparable to the mass-step of SN Ia hosts.

\begin{table*}
\centering
\begin{tabular}{c|c|c}
\hline
 LWA-to-MWA conversions              & $\Delta {\rm HR}_{\rm mass}$ & $\Delta {\rm HR}_{\rm age}$ \\ \hline
 LWA-to-MWA 1                        & $0.1126 \pm 0.0009$ & $0.1968 \pm 0.0007$ \\
 LWA-to-MWA 2                        & $0.0895 \pm 0.0008$ & $0.1589 \pm 0.0008$ \\
 LWA-to-MWA 1 w/ 1.0 Gyr age scatter & $0.1156 \pm 0.0008$ & $0.1618 \pm 0.0007$ \\
 LWA-to-MWA 1 w/ conversion error    & $0.1005 \pm 0.0011$ & $0.2427 \pm 0.0008$ \\
 LWA-to-MWA 2 w/ conversion eeror    & $0.0981 \pm 0.0011$ & $0.2391 \pm 0.0009$ \\
 G05 LWA-to-MWA 1                    & $0.1438 \pm 0.0009$ & $0.2278 \pm 0.0008$ \\
 G05 LWA-to-MWA 2                    & $0.1124 \pm 0.0009$ & $0.1959 \pm 0.0008$ \\ \hline
\end{tabular}
\caption{
Simulated $\Delta {\rm HR}$ for various LWA-to-MWA conversions.
The host mass and host age thresholds used for $\Delta {\rm HR}$ are consistent with those in Figure~\ref{f3}.}
\label{t1}
\end{table*}

The upcoming surveys, such as those from the James Webb Space Telescope, are anticipated to detect a larger number of SNe Ia at high redshifts. 
If our analysis is correct, the host mass-step observed in the high-$z$ SN Ia sample would show a smaller magnitude difference because the progenitor age range is reduced as the age of the red sequence becomes younger (see Figure~6 of \citealt{2022MNRAS.tmp.2647L}).
In this respect, it is interesting to note that a comparison between the local sample \citep{2010ApJ...715..743K} and the SNLS sample at $z \sim 1.0$ \citep{2011ApJ...737..102S} already indicates a hint of this trend. 
Also, measuring reliable population ages for a large sample of host galaxies, both at high-$z$ and low-$z$, would help to further elucidate the root cause of the host mass-step.

\acknowledgments
We thank the anonymous referee for a number of helpful comments.
We acknowledge support from the National Research Foundation of Korea to the Center for Galaxy Evolution Research (2022R1A6A1A03053472). 
Y.-W.L. and S.-J.Y. acknowledge support from the Mid-career Researcher Program (2022R1A2C3002992, 2019R1A2C3006242) through the National Research Foundation of Korea.

\section{Appendix}

Here we show the effect of the applied AMRs and the age scatter on the projection simulations. 
The top panels of Figure~\ref{f1a} display the rescaled AMRs obtained from the LWA-to-MWA conversion 1 and 2, as well as conversion 2’s 13.6th and 84.6th percentile conversions, along with the case of the G05 AMR. 
Although the rescaled AMR obtained through the LWA-to-MWA conversion 2 demonstrates a smaller age difference between low- and high-mass galaxies, the resulting AMR (solid line) maintains the `S'-shape similar to the rescaled AMR based on the LWA-to-MWA conversion 1 (short dashed line). 
Our LWA-to-MWA conversion 2 resulted in the narrower age distribution (solid red histogram in panel ($d$)) compared to that from the LWA-to-MWA conversion 1 (short dashed black histogram in the same panel), with a mean age difference of 3.90~Gyr between the two groups. 
However, the distribution's $D$-value of 2.20 and the comparable fraction ($42.6 \pm 3.3$\% : $57.4\pm 3.3$\%) still indicates strong bimodality in age. 
In order to assess the impact of extreme outliers on the LWA-to-MWA conversion, we examine two additional scenarios involving age conversion errors at the 13.6th and 86.4th percentiles. 
Projected age distributions with these errors also result in bimodal distributions, with D-values of 2.11 and 2.00 for the LWA-to-MWA conversion 1 and 2, respectively. 
The proportions of old age groups for the LWA-to-MWA conversion 1 and 2 are $84.1 \pm 2.0$\% and $79.2 \pm 1.4$\%, respectively. 
This reaffirms the significance of nonlinear relationships in understanding the bimodal age distribution of galaxies, regardless of the age conversion errors.

We perform additional projection simulations based on the G05 AMR. 
Both AMRs based on the LWA-to-MWA conversion 1 and 2 for the G05 AMRs exhibit clearly bimodal age distributions with even larger age differences of 6.15 and 5.19~Gyr, respectively. 
The $D$-values of the distributions are 2.88 and 2.53 with the fraction of the old age groups of $59.1\pm1.9$\% and $65.6\pm1.8$\%, respectively. 
On the other hand, the red dashed histogram in Figure~\ref{f1a}($d$) is for the simulation based on the LWA-to-MWA conversion 1 that employs the age scatter of 1.0~Gyr. 
The overall age distribution shows clear bimodality in comparison to Figure~\ref{f2}($d$), due to the reduced age scatter (in contrast to the age scatter employed throughout the manuscript). 
It is important to note that the quantitative age difference is strongly influenced by the employed stellar population models, but nonetheless, bimodality persists in the qualitative morphology of the age distributions.
We have summarized the resulting $\Delta {\rm HR}$ values in Table~\ref{t1} for various assumptions regarding LWA-to-MWA conversion and associated errors. 
Although minor differences exist among the different conversions and the applied errors, the simulated $\Delta {\rm HR}_{\rm mass}$ values consistently yield approximately $\sim 0.1$~mag, irrespective of the chosen conversions or conversion errors.
This result further confirms the close relationship between bimodal age distributions and the host mass-steps. 


\end{document}